\documentclass[conference,letterpaper]{IEEEtran}
\IEEEoverridecommandlockouts
\usepackage{cite}
\usepackage{amsmath,amssymb,amsfonts}
\usepackage{algorithmic}
\usepackage{graphicx}
\usepackage{textcomp}
\usepackage{balance}
\usepackage{xcolor}
\usepackage{siunitx}
\usepackage[caption=false,font=footnotesize]{subfig}
\usepackage{cleveref}
\def\BibTeX{{\rm B\kern-.05em{\sc i\kern-.025em b}\kern-.08em
    T\kern-.1667em\lower.7ex\hbox{E}\kern-.125emX}}

\crefname{figure}{Fig.}{Figs.}

\begin{document}

\title{Comparison of Indicators of Location Homophily Using Twitter Follow Graph}

\author{%
  \IEEEauthorblockN{\small Shiori Hironaka \quad Mitsuo Yoshida \quad Kyoji Umemura}
  \IEEEauthorblockA{\small \textit{Department of Computer Science and Engineering}\\
  \textit{Toyohashi University of Technology}\\
  Aichi, Japan\\
  s143369@edu.tut.ac.jp, yoshida@cs.tut.ac.jp, umemura@tut.jp}
}


\maketitle

\begin{abstract}
Location homophily is a tendency of Twitter users whose followers tend to be in the same or nearby areas.
Intuitively, although users with a higher number of follower relationships might have negative homophily indicators, it is worth consulting actual Twitter data.
Moreover, there may be certain functions regarding the numbers of friends and followers that are more directly correlated to the homophily.
In this study, the ratio of the number of friends to the number of followers is shown to be a more effective negative indicator of homophily, and the results for 10 different countries are verified.
\end{abstract}

\vspace{\baselineskip}
\renewcommand\IEEEkeywordsname{Keywords}

\section{Introduction}

The homophily of a person is a valuable concept that has been extensively studied in the field of social science~\cite{McPherson2001,Block2014}.
In this paper, location homophily indicates that people who share a location of activity tend to be friends.
Location homophily is an interesting subject in online social graphs, where the relationship between users is not necessarily physical.
This subject is closely related to the majority of research on home location estimation using online social graphs~\cite{Jurgens2015,Luo2020,Zheng2018}, where the location homophily is the assumption used for the estimation.

In a home location estimation, it was reported that if nodes of large degrees are removed, the estimation becomes more effective~\cite{Rahimi2015}.
This suggests that such users tend to show less location homophily, and that the numbers of followers (\#followers) and friends (\#friends) are negative indicators of homophily.
It is interesting to examine whether \#friends or \#followers are better indicators of homophily, and in early research on homophily on Twitter~\cite{De2011}, the ``ego ratio,'' the ratio of \#followers to \#friends, was used as an indicator of homophily.
However, comparisons of this ratio with \#followers or \#friends have yet to be reported.
It is also interesting to examine whether this relationship is consistent, even in different countries.

In this study, the homophily of a set of users is measured based on the accuracy of home location estimation using the followers' locations, as determined through the majority vote.
To judge whether a particular attribute such as \#followers, \#friends, or their ratio (\#friends/\#followers) is an effective indicator of homophily, users are selected by the attribute and seek the best filter threshold to maximize the accuracy of the home location estimation.
The accuracy of a good negative indicator of homophily improves through the removal of high values of the attribute.
To make the analysis more robust, the same operation was applied to 10 different countries.
The results show that \#friends/\#followers is a universally effective negative indicator of homophily.

Our conclusions are as follows:
\begin{itemize}
  \item The follow ratio (\#friends/\#followers) is a more stable value among the 10 countries than \#friends and \#followers.
  \item The follow ratio is a robust negative indicator of the location homophily.
  \item Among \#friends, \#followers, and \#friends/\#followers, \#friends/\#followers is the strongest indicator of location homophily.
\end{itemize}

\section{Related Work}

Homophily has been extensively studied in social science~\cite{McPherson2001}.
The homophily of race, ethnicity, gender, and location, among other factors, as well as their combinations, has been investigated~\cite{Block2014}.
Homophily has also been observed in online relationships, and its effect has been used in many studies, such as recommendation systems~\cite{Wang2017} and attribute estimation~\cite{Jurgens2015,Li2014}.

The relationship of homophily and a tie formation is focused on in~\cite{De2011}, where
the authors introduced the ``ego ratio,'' the ratio of \#followers to \#friends, and compared it with the homophily measured through such attributes as location, gender, ethnicity, or sentiment expression.
The authors also showed that the strength of the homophily differed depending on the type when the users were divided into three types based on their ego ratio.
The higher the ego ratio, the stronger the location homophily was observed.
Although the ego ratio is a function of \#friends and \#followers, the relationship between the location homophily and \#friends and the relationship between location homophily and \#followers are not presented in~\cite{De2011}.
It is unclear which is the best indicator of the location homophily.
In addition, the location information used in~\cite{De2011} is at the country level, which is larger than the data used.

Hironaka et al.~\cite{Hironaka2021} analyzed the profile attributes that contribute to the performance of home location estimation in 10 countries.
Many profile attributes, including the number of friends and followers, were used in the analysis.
They found that the only attribute that was effective in all 10 countries was the follow ratio.
They focused only on whether an attribute contributed to the estimation performance and not on the degree of improvement.

In this study, we compare the degree to which \#friends, \#followers, and the follow ratio indicate the location homophily using the location estimation.
As a result, we found that the follow ratio is the most robustly negative indicator of the location homophily.

\section{Data}

We applied social graph data, home location data, and attribute data for this experiment.
For this, we used the datasets from a previous study~\cite{Hironaka2021},
which consist of social graph data, home locations, and user attributes in 10 countries.

\subsection{Dataset Description}

Please see~\cite{Hironaka2021} for details of the dataset creation.
In this dataset, the home location data were created using geo-tagged tweets collected in 2019, and social graph data were constructed using the mutual friend relationships between users.
The user attributes, number of friends (\#friends), and number of followers (\#followers) were extracted from the profile data attached to the newest tweets collected.
Because it is natural to assume that every user also follows oneself, the follow ratio (\#friends/\#followers) is calculated as $(\textrm{\#friends} + 1) / (\textrm{\#followers} + 1)$.

\Cref{tb:graph-properties} shows the statistics of the datasets, where $|V|$ is the number of users, $|E|$ is the number of edges (mutual friend relationships), $|I|$ is the number of isolated users (users with no mutual friends), $K_\mathrm{out}$ is the average number of mutual friends, $S_\mathrm{out}$ is the variance of the number of mutual friends, and $M_\mathrm{out}$ is the median of the number of mutual friends.

\begin{table}[tp]
\centering
\caption{Basic statistics of the datasets.}
\label{tb:graph-properties}
  \resizebox{\columnwidth}{!}{%
\begin{tabular}{lrrrrrr} \hline
Country & $|V|$    & $|I|$    & $|E|$      & $K_\mathrm{out}$  & $S_\mathrm{out}$   & $M_\mathrm{out}$ \\ \hline
  United States  & \num{1748492} & \num{206671} & \num{80973364} & 41.42 & 164.91 & 14   \\
  Brazil         & \num{591526}  & \num{93794}  & \num{35659270} & 52.03 & 133.48 & 18   \\
  United Kingdom & \num{395492}  & \num{50172}  & \num{14835714} & 33.29 & 86.88  & 11   \\
  Japan          & \num{325846}  & \num{70033}  & \num{9305522}  & 23.51 & 72.68  & 6    \\
  Philippines    & \num{207758}  & \num{71518}  & \num{5756312}  & 20.61 & 63.94  & 7    \\
  Turkey         & \num{188928}  & \num{47822}  & \num{5857770}  & 24.74 & 179.36 & 5    \\
  Indonesia      & \num{197038}  & \num{57486}  & \num{2788522}  & 10.96 & 32.56  & 4    \\
  India          & \num{122686}  & \num{98031}  & \num{2318458}  & 10.50 & 58.62  & 1    \\
  Mexico         & \num{165315}  & \num{35703}  & \num{3933724}  & 19.57 & 69.72  & 5    \\
  Saudi Arabia   & \num{128700}  & \num{46179}  & \num{2590364}  & 14.81 & 68.35  & 3   \\ \hline
\end{tabular}%
}
\end{table}

\subsection{Statistics and Correlations of Attributes}

\begin{figure}
  \centering
  \subfloat[Number of friends]{%
    \includegraphics[width=\linewidth]{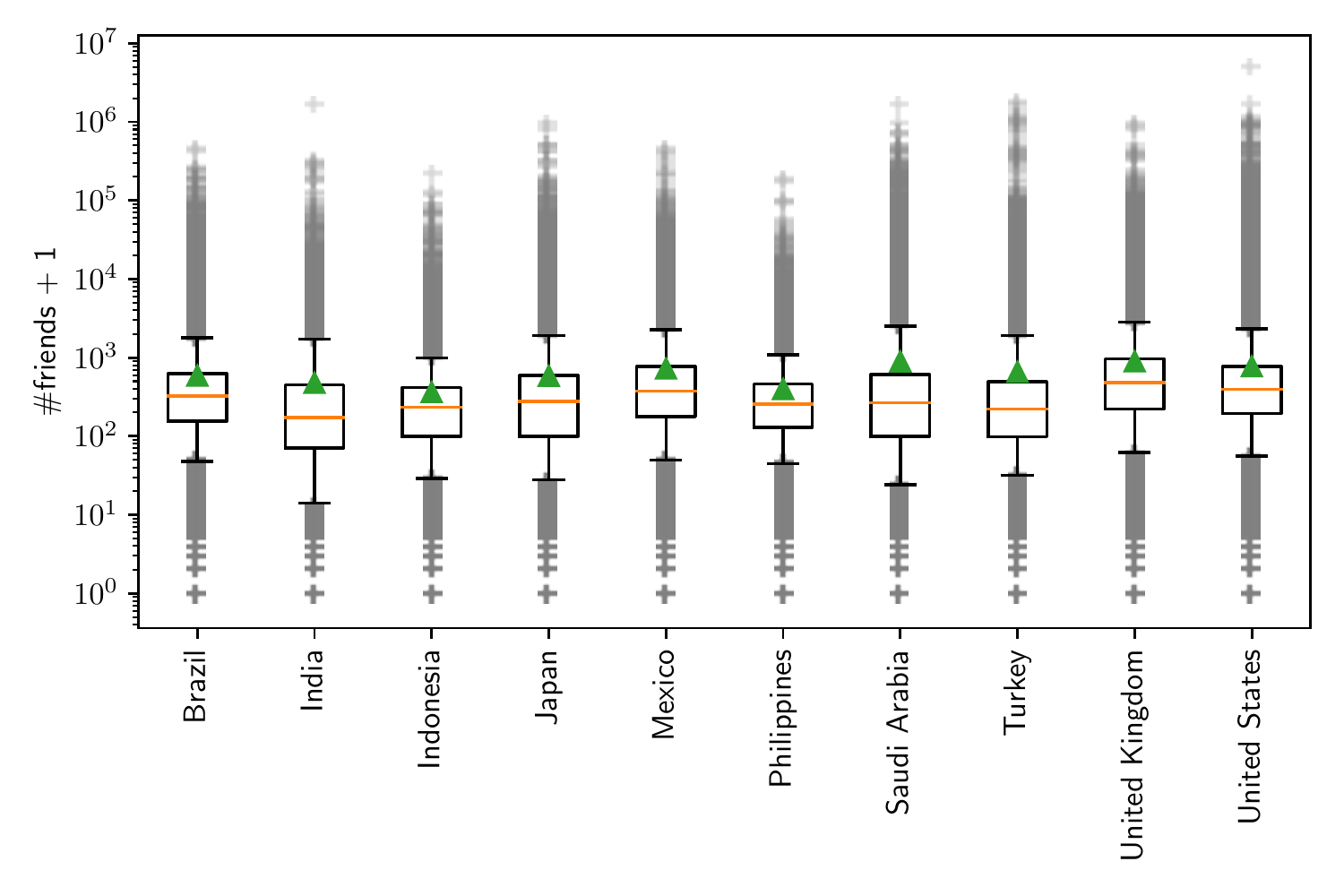}%
  }
  \\
  \subfloat[Number of followers]{%
    \includegraphics[width=\linewidth]{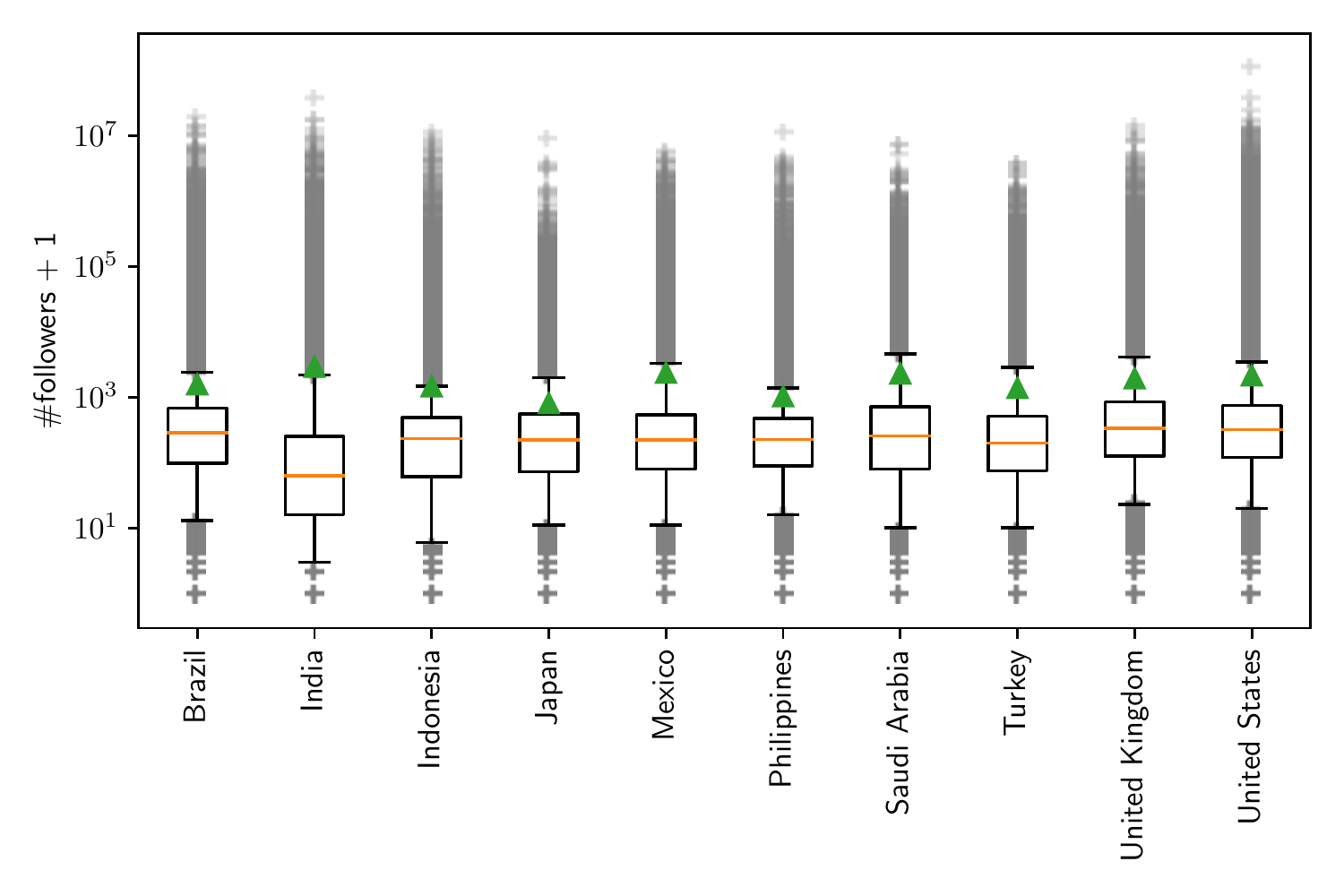}%
  }
  \\
  \subfloat[Follow ratio]{%
    \includegraphics[width=\linewidth]{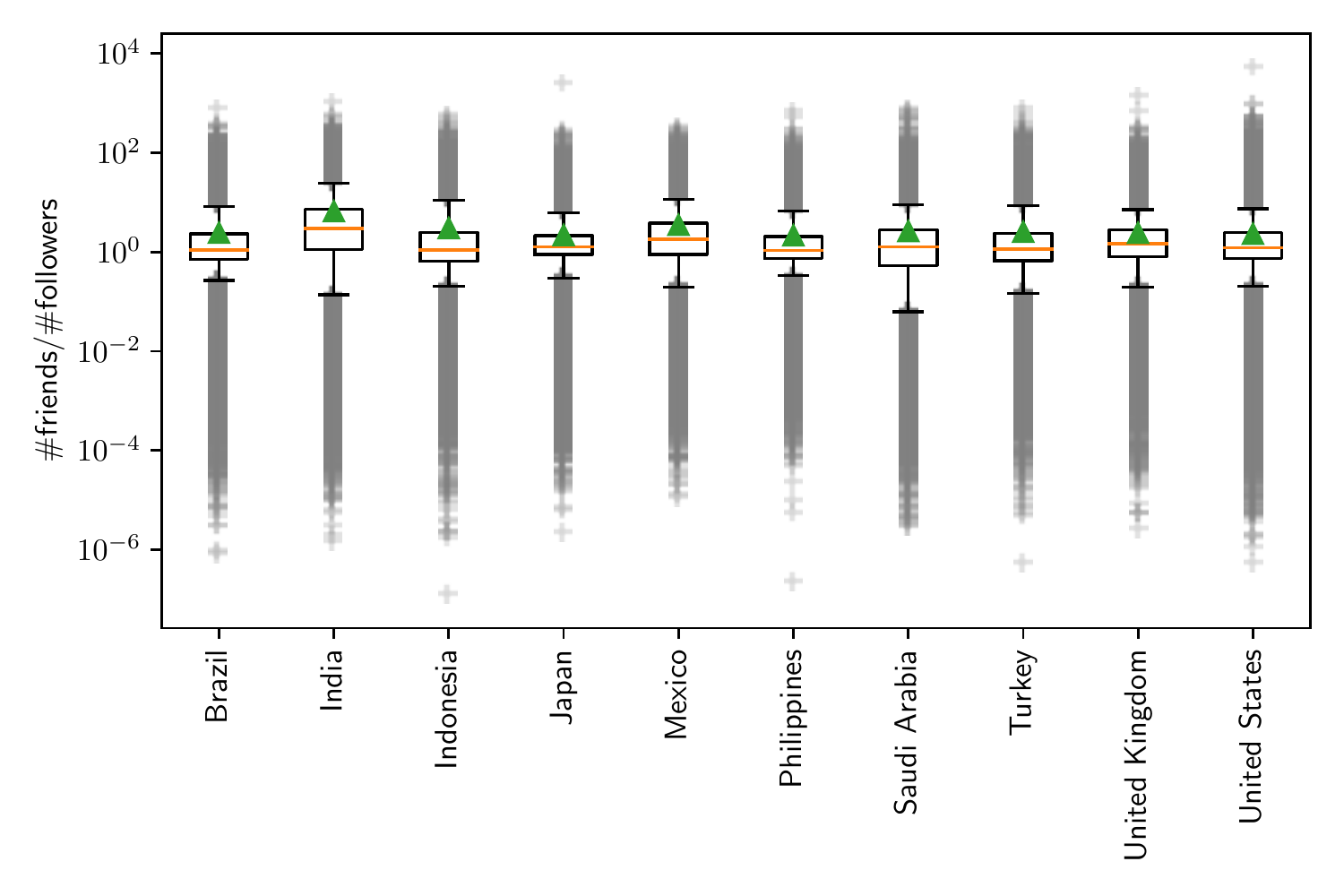}%
  }
  \caption{Box plot of the attributes in each country. There are little differences in the follow ratios among countries when compared with the differences in the numbers of friends or followers.}
  \label{fig:dist}
\end{figure}

We compute the statistics of the user attributes as follows.
\Cref{fig:dist} shows boxplots of the number of friends, the number of followers, and the follow ratio.
The box indicates the range of values from the 25th to 75th percentiles.
The whiskers indicate the range of values from the 5th to 95th percentiles.
The middle line in the box represents the median value, and the triangular dot represents the mean value.
Outliers are plotted as a plus mark (``+'').
In \cref{fig:dist}, there are differences in the number of friends and followers between countries.
By contrast, the difference in the follow ratio between countries is small.

We compute Spearman's rank correlation coefficient between attributes.
\Cref{tb:corr} shows the correlations for all combinations between the number of friends, the number of followers, and the follow ratio.
There are strong correlations ($corr>0.5$) between the number of friends and the number of followers and between the number of followers and the follow ratio.
In addition, there is an extremely weak correlation between the number of friends and the follow ratio ($|corr|<0.2$).
Similar trends are obtained for all 10 countries.
Consequently, we can compare the relative differences of the attributes within countries.

\begin{table}
  \centering
\caption{Correlations between the number of friends, the number of followers, and the follow ratio. The follow ratio is extremely weakly correlated with the number of friends.}\label{tb:corr}
\begin{tabular}{lSSS}
\hline
  Country & {friends--followers} & {friends--ratio} & {followers--ratio} \\ \hline
United States & 0.69  & 0.03  & -0.64  \\
Brazil & 0.72  & -0.06  & -0.68  \\
United Kingdom & 0.70  & 0.03  & -0.64  \\
Japan & 0.80  & -0.04  & -0.55  \\
Philippines & 0.72  & -0.02  & -0.65  \\
Turkey & 0.64  & 0.11  & -0.62  \\
Indonesia & 0.66  & -0.07  & -0.73  \\
India & 0.65  & 0.03  & -0.70  \\
Mexico & 0.64  & 0.06  & -0.67  \\
Saudi Arabia & 0.56  & 0.19  & -0.64  \\
\hline
\end{tabular}
\end{table}

\section{Experimental Method}

We conducted an experiment to find the most effective attribute for capturing the location homophily.
The strength of the location homophily was measured based on the performance of the home location estimation using a social graph.

The experiment proceeded as follows.
First, the performance of each user's home location was evaluated using a leave-one-out cross-validation without filtering.
Next, the location estimation was evaluated after filtering the estimation targets using each attribute.
The estimation performances were then compared.

\subsection{Location Estimation Method}

The location estimation task is to estimate the true label $Y_U$ of users $U$, given a social graph $G(V, E)$, the known location labels $Y_L$ of the users $V_L (V_L \subset V)$, and the target users $U$.
The task is formulated as follows:
\begin{equation}
  \hat{Y}_U = \mathrm{Infer}(V, E, Y_L),
\end{equation}
where $\hat{Y}_U$ denotes the estimated labels of the users $U$.

In this study, we employ the method proposed by Davis et al.~\cite{Davis2011}, which showed a high estimation accuracy in comparison to Jurgens et al.~\cite{Jurgens2015}.
This estimation method assumes that the majority of the connected users on a social graph have the same location label and estimates the labels using a majority vote of the labels held by the adjacent nodes.

\subsection{Filtering Method}

A filter divides the user set into two parts at a threshold using the number of friends, the number of followers, and the follow ratio.
We use two types of filters: a HighCut filter that filters users with values higher than the threshold $\theta$, and a LowCut filter that filters users with values lower than the threshold $\theta$.

The filters are expressed as follows:
\begin{align}
  \mathrm{HighCut}(V, a) = \{ u | u \in V, a_u < \theta \}, \\
  \mathrm{LowCut}(V, a) = \{ u | u \in V, a_u > \theta \},
\end{align}
where $V$ is a set of users, $a$ is one of three attributes, and $a_u$ is the attribute value of user $u$.

The filter works well if there is a linear relationship between the location homophily and the attribute around the threshold.
We consider that the number of friends and followers are characterized by parts with high values, for which we apply the HighCut filter.
We apply the two filters to the follow ratio because the follow ratio is characterized by two parts that are greater and less than 1, respectively.
Only one filter is applied at a time.

Each filter has the threshold $\theta$ as a parameter.
By dividing the value of each attribute from the minimum to maximum values into 100 parts at equal intervals on a logarithmic scale, we find the threshold that maximizes the estimation accuracy under a coverage of greater than 0.3.
We computed the results for all test data at all thresholds and reported the point at which the estimation accuracy was maximized.

\subsection{Evaluation}

The estimation was measured in terms of accuracy and coverage.
The error distance was not used for the evaluation because the area size differs between countries.
The accuracy is the percentage of correctly estimated labels.
Coverage is the percentage of users for whom we can estimate any labels.
For this experiment, the coverage is the percentage of users who have mutual friends that can be used for the estimation.

The metrics are expressed as follows:
\begin{align}
  \mathrm{Accuracy}(U) = \frac{ |\{ u \mid u \in U, y_u = \hat{y}_u \}| }{ | \{ u \mid u \in U, |N_u| > 0 \} | }, \\
  \mathrm{Coverage}(U) = \frac{|\{ u \mid u \in U, |N_u| > 0 \}|}{|V|},
\end{align}
where $y_u$ is the true label of user $u$, $\hat{y}_u$ is an estimated label of user $u$, $U$ is a set of target users of the estimation (the filtered users, if a filter is applied), $N_u$ is an adjacent node of user $u$ (mutual friends of $u$), and $V$ is a set of users in the dataset.

\subsection{Significance Testing}

To judge whether the accuracy was improved by the filter applied, we statistically tested the accuracy as follows.
Let the null hypothesis be $H_0$ and the alternative hypothesis be $H_1$, where the accuracy without the filter is $p_1$ and the accuracy with the filter applied is $p_2$.
\begin{align}
H_0: p_1 = p_2 \Rightarrow p_2 - p_1 = 0 \\
H_1: p_1 < p_2 \Rightarrow p_2 - p_1 > 0
\end{align}
Letting $x_1$ and $x_2$ denote the number of correct answers, and $n_1$ and $n_2$ denote the number of answers, the accuracies $p_1$ and $p_2$ are estimated as $\hat{p}_1 = x_1 / n_1$ and $\hat{p}_2 = x_2 / n_2$, respectively.
When $n$ users are randomly sampled, the distribution of accuracy $p$ follows a normal distribution with mean $\hat{p}$ and variance $\hat{p}(1-\hat{p})/n$.
The parameters of the sum of the normal distributions are easily obtained.
\begin{align}
  p_2 - p_1 &\sim N(\hat{p}_2 - \hat{p}_1, \frac{\hat{p}_1(1-\hat{p}_1)}{n_1} + \frac{\hat{p}_2(1-\hat{p}_2)}{n_2})
\end{align}

The confidence interval of $p_2 - p_1$ is expressed through the following equation:
If this equation does not contain zero, we consider $p_1$ to be significantly greater than $p_2$.
Here, $z_\alpha$ is the z-score of the standard normal distribution at point $\alpha$.
During the experiment, we use $\alpha = 0.05$.
\begin{multline}
\hat{p_2} - \hat{p_1} - z_\alpha \sqrt{\frac{\hat{p}_1(1-\hat{p}_1)}{n_1} + \frac{\hat{p}_2(1-\hat{p}_2)}{n_2}} \leq p_2 - p_1 \\
  \leq \hat{p_2} - \hat{p_1} + z_\alpha \sqrt{\frac{\hat{p}_1(1-\hat{p}_1)}{n_1} + \frac{\hat{p}_2(1-\hat{p}_2)}{n_2}}
\end{multline}

\section{Results}

\Cref{tb:result} shows the results of the experiment.
If the accuracy is significantly improved compared to that without filtering, the value is marked with an asterisk ($^*$).
First, the coverage without filtering of each country must be discussed.
In every country, more than half of Twitter users have at least one mutual friend; thus, the coverage is sufficient to discuss the effect of the selection through filtering.
The accuracy without filtering of each country should then be examined.
The accuracy is within the range of 0.186 to 0.658, and the values vary by country.
There may be many reasons for this, such as culture, popularity, and population density.
This implies that comparing the absolute value of accuracy between countries is difficult to analyze.
Instead, the ratio of accuracy with filtering to accuracy without filtering is applied to reflect the tendency of the attributes.
Therefore, even if the accuracy is approximately 0.1, it should be possible to discuss the attributes.

If the filter does not work, the corresponding coverage becomes the same or very close to that without filtering.
The filter works well when the corresponding coverage becomes approximately half of the coverage without filtering, and the accuracy increases.
The value of the optimum threshold is meaningful only when the filter works well.
The threshold of the filter is determined such that the corresponding coverage is greater than 0.3, and the corresponding accuracy is the highest.

For \cref{tb:result}, which details the data regarding \#friends, nine countries (all excluding Japan) show an improvement.
The \#friends filter works in Indonesia, India, Mexico, and Saudi Arabia (4 out of 10 countries) because of their coverage values.
For \#followers, the filter works well in Saudi Arabia only.
The LowCut filter can be used for the follow ratio in Brazil, Japan, and the Philippines, but does not work well even in these countries because of the coverage values.

The HighCut filter works well for the follow ratio in all 10 countries.
Each coverage value becomes approximately half of the coverage without filtering, and each accuracy value is considerably better than that without filtering.
It should also be noted that the best thresholds are close to 1.0, except for India, for which the \#friends filter works extremely well.

\begin{table*}
  \centering
\caption{Accuracy and coverage results for each filter and country. The HighCut filter works well for the follow ratio in all 10 countries.}\label{tb:result}
\begin{tabular}{llcrSS}
\hline
  Country & Attribute & Filter & Threshold & {Accuracy} & {Coverage} \\
\hline
  United States & \#friends & HighCut & 1452  & 0.359$^*$  & 0.792  \\
& \#followers & HighCut & 108004699  & 0.353  & 0.894  \\
& \#friends/\#followers & HighCut & 0.923  & 0.413$^*$  & 0.333  \\
& \#friends/\#followers & LowCut & 0.000  & 0.353  & 0.894  \\
&  & none &  & 0.353  & 0.894  \\ \hline
Brazil & \#friends & HighCut & 800  & 0.666$^*$  & 0.695  \\
& \#followers & HighCut & 1720  & 0.656$^*$  & 0.789  \\
& \#friends/\#followers & HighCut & 0.979  & 0.704$^*$  & 0.393  \\
& \#friends/\#followers & LowCut & 0.227  & 0.650$^*$  & 0.828  \\
&  & none &  & 0.648  & 0.863  \\ \hline
United Kingdom & \#friends & HighCut & 3198  & 0.305  & 0.849  \\
& \#followers & HighCut & 12327  & 0.304  & 0.872  \\
& \#friends/\#followers & HighCut & 1.166  & 0.327$^*$  & 0.370  \\
& \#friends/\#followers & LowCut & 0.024  & 0.303  & 0.881  \\
&  & none &  & 0.303  & 0.887  \\ \hline
Japan & \#friends & HighCut & 38100  & 0.186  & 0.823  \\
& \#followers & HighCut & 36456  & 0.186  & 0.822  \\
& \#friends/\#followers & HighCut & 1.202  & 0.215$^*$  & 0.408  \\
& \#friends/\#followers & LowCut & 0.000  & 0.186  & 0.823  \\
&  & none &  & 0.186  & 0.823  \\ \hline
Philippines & \#friends & HighCut & 580  & 0.487$^*$  & 0.598  \\
& \#followers & HighCut & 1339  & 0.482$^*$  & 0.696  \\
& \#friends/\#followers & HighCut & 1.101  & 0.518$^*$  & 0.408  \\
& \#friends/\#followers & LowCut & 0.365  & 0.480$^*$  & 0.697  \\
&  & none &  & 0.476  & 0.744  \\ \hline
Turkey & \#friends & HighCut & 596  & 0.502$^*$  & 0.610  \\
& \#followers & HighCut & 3774006  & 0.496  & 0.798  \\
& \#friends/\#followers & HighCut & 1.059  & 0.518$^*$  & 0.397  \\
& \#friends/\#followers & LowCut & 0.000  & 0.496  & 0.798  \\
&  & none &  & 0.496  & 0.798  \\ \hline
Indonesia & \#friends & HighCut & 211  & 0.154$^*$  & 0.307  \\
& \#followers & HighCut & 1339  & 0.144  & 0.723  \\
& \#friends/\#followers & HighCut & 0.845  & 0.164$^*$  & 0.307  \\
& \#friends/\#followers & LowCut & 0.112  & 0.143  & 0.754  \\
&  & none &  & 0.143  & 0.774  \\ \hline
India & \#friends & HighCut & 366  & 0.298$^*$  & 0.305  \\
& \#followers & HighCut & 36043189  & 0.283  & 0.556  \\
& \#friends/\#followers & HighCut & 2.566  & 0.325$^*$  & 0.312  \\
& \#friends/\#followers & LowCut & 0.000  & 0.283  & 0.556  \\
&  & none &  & 0.283  & 0.556  \\ \hline
Mexico & \#friends & HighCut & 550  & 0.375$^*$  & 0.489  \\
& \#followers & HighCut & 14209  & 0.368  & 0.807  \\
& \#friends/\#followers & HighCut & 1.218  & 0.422$^*$  & 0.321  \\
& \#friends/\#followers & LowCut & 0.013  & 0.367  & 0.815  \\
&  & none &  & 0.367  & 0.822  \\ \hline
Saudi Arabia & \#friends & HighCut & 277  & 0.623$^*$  & 0.312  \\
& \#followers & HighCut & 793  & 0.593$^*$  & 0.528  \\
& \#friends/\#followers & HighCut & 1.030  & 0.627$^*$  & 0.347  \\
& \#friends/\#followers & LowCut & 0.000  & 0.580  & 0.736  \\
&  & none &  & 0.580  & 0.736  \\
\hline
\end{tabular}
\end{table*}

\section{Discussion}

Cases in which user A is followed by user B because user A started following user B is rather common.
This observation is important for understanding the relationship between \#friends and \#friends/\#followers.
The relationship between acquaintances and location homophily also plays an important role in interpreting the results.

\subsection{Importance of \#friends and \#friends/\#followers}

It is common practice to use \#friends and \#followers to estimate the user characteristics.
In particular, users who have large \#followers are said to be influential~\cite{Kwak2010}.
The correlation analysis between \#friends, \#followers, and \#friends/\#followers suggests a different way to describe users.
 Because \#friends/\#followers is calculated as $(\textrm{\#friends} + 1) / (\textrm{\#followers} +1)$, one may think that the correlation between \#friends and \#friends/\#followers would be positive; however, this is not the case.
It is very interesting to note that there is no correlation between \#friends and the follow ratio for all countries.
This implies that \#friends and the follow ratio are natural attributes that can be used to characterize users, and \#followers can be predicted by the other two attributes.
In other words, although a user may follow as many users as they like, the probability of being followed back does not depend on the number of friends (\#friends).

\subsection{What does \#friends indicate?}

One reasonable interpretation of \#friends is the degree of activity.
If an active user naturally follows other users, then \#friends will be proportional to the user's activity.
If \#friends more than 1000, the neighbors of user should be distributed over a wide range of area.
This explains the results indicating that the best threshold for \#friends is large, and only a small percentage of users should be filtered.

\subsection{What does the follow ratio indicate?}

One reasonable interpretation of the follow ratio is the ratio of non-acquainted neighbors to acquainted neighbors, where user A's acquainted neighbors are the users mutually following user A.
A user with a relatively large follow ratio may try to establish a new relation by following others.
If the majority of follow actions from others are triggered by the user's follow actions, then \#friends should be greater than \#followers.
In this case, the follow ratio becomes greater than 1.
A user with a relatively small follow ratio may try to follow users who are already acquainted.
In this case, the majority of follow actions from others are not triggered by the follow action.
If a user who has many followers without following others follows an acquainted user, the user is likely to follow back.
In this case, the follow ratio becomes less than 1.
Interestingly, the follow ratio is the best negative indicator of location homophily for all 10 countries.

\subsection{Why is the follow ratio more effective than \#friends?}

If the majority of one user's neighbors are acquainted, the location homophily should be large.
As previously discussed, the follow ratio is directly related to the ratio of acquaintances among neighbors, whereas \#friends is related to the ratio of acquaintances among neighbors only for rare and uncommon users.
Although it is difficult to judge whether one user is really acquainted with others in practice, the location homophily can be estimated.

\section{Conclusion}

In this study, the location homophily of a set of users was defined as the accuracy of the home location estimation.
The number of friends, number of followers, and follow ratio were selected as attributes.
By filtering using these attributes, the accuracy and coverage of the home location estimation were obtained.
The same operation was applied to 10 different countries.
As a result, the follow ratio was shown to be the most effective attribute for capturing the location homophily.

\balance
\bibliographystyle{IEEEtran}
\bibliography{./references}

\begin{thebibliography}{10}
\providecommand{\url}[1]{#1}
\csname url@samestyle\endcsname
\providecommand{\newblock}{\relax}
\providecommand{\bibinfo}[2]{#2}
\providecommand{\BIBentrySTDinterwordspacing}{\spaceskip=0pt\relax}
\providecommand{\BIBentryALTinterwordstretchfactor}{4}
\providecommand{\BIBentryALTinterwordspacing}{\spaceskip=\fontdimen2\font plus
\BIBentryALTinterwordstretchfactor\fontdimen3\font minus
  \fontdimen4\font\relax}
\providecommand{\BIBforeignlanguage}[2]{{%
\expandafter\ifx\csname l@#1\endcsname\relax
\typeout{** WARNING: IEEEtran.bst: No hyphenation pattern has been}%
\typeout{** loaded for the language `#1'. Using the pattern for}%
\typeout{** the default language instead.}%
\else
\language=\csname l@#1\endcsname
\fi
#2}}
\providecommand{\BIBdecl}{\relax}
\BIBdecl

\bibitem{McPherson2001}
M.~McPherson, L.~Smith-Lovin, and J.~M. Cook, ``Birds of a feather: Homophily
  in social networks,'' \emph{Annual Review of Sociology}, vol.~27, no.~1, pp.
  415--444, 2001.

\bibitem{Block2014}
P.~Block and T.~Grund, ``Multidimensional homophily in friendship networks,''
  \emph{Network Science}, vol.~2, no.~2, pp. 189--212, 2014.

\bibitem{Jurgens2015}
D.~Jurgens, T.~Finethy, J.~Mccorriston, Y.~T. Xu, and D.~Ruths, ``Geolocation
  prediction in {Twitter} using social networks: A critical analysis and review
  of current practice,'' in \emph{Proceedings of the 9th International AAAI
  Conference on Web and Social Media}, 2015, pp. 188--197.

\bibitem{Luo2020}
X.~Luo, Y.~Qiao, C.~Li, J.~Ma, and Y.~Liu, ``An overview of microblog user
  geolocation methods,'' \emph{Information Processing \& Management}, vol.~57,
  no.~6, p. 102375, 2020.

\bibitem{Zheng2018}
X.~Zheng, J.~Han, and A.~Sun, ``A survey of location prediction on {Twitter},''
  \emph{IEEE Transactions on Knowledge and Data Engineering}, vol.~30, no.~9,
  pp. 1652--1671, 2018.

\bibitem{Rahimi2015}
A.~Rahimi, T.~Cohn, and T.~Baldwin, ``Twitter user geolocation using a unified
  text and network prediction model,'' in \emph{Proceedings of the 53rd Annual
  Meeting of the Association for Computational Linguistics and the 7th
  International Joint Conference on Natural Language Processing}, 2015, pp.
  630--636.

\bibitem{De2011}
M.~{De Choudhury}, ``Tie formation on {Twitter}: Homophily and structure of
  egocentric networks,'' in \emph{Proceedings of the 2011 IEEE Third
  International Conference on Privacy, Security, Risk and Trust and 2011 IEEE
  Third International Conference on Social Computing}, 2011, pp. 465--470.

\bibitem{Wang2017}
X.~Wang, S.~C. Hoi, M.~Ester, J.~Bu, and C.~Chen, ``Learning personalized
  preference of strong and weak ties for social recommendation,'' in
  \emph{Proceedings of the 26th International Conference on World Wide Web},
  2017, pp. 1601--1610.

\bibitem{Li2014}
R.~Li, C.~Wang, and K.~C.-C. Chang, ``User profiling in an ego network:
  Co-profiling attributes and relationships,'' in \emph{Proceedings of the 23rd
  International Conference on World Wide Web}, 2014, pp. 819--830.

\bibitem{Hironaka2021}
S.~Hironaka, M.~Yoshida, and K.~Umemura, ``Cross-country analysis of user
  profiles for graph-based location estimation,'' \emph{IEEE Access}, 2021, in
  press.

\bibitem{Davis2011}
C.~A. {Davis Jr.}, G.~L. Pappa, D.~R.~R. de~Oliveira, and F.~{de L. Arcanjo},
  ``Inferring the location of {Twitter} messages based on user relationships,''
  \emph{Transactions in GIS}, vol.~15, no.~6, pp. 735--751, 2011.

\bibitem{Kwak2010}
H.~Kwak, C.~Lee, H.~Park, and S.~Moon, ``What is {Twitter}, a social network or
  a news media?'' in \emph{Proceedings of the 19th International Conference on
  World Wide Web}, 2010, pp. 591--600.

\end{thebibliography}

\end{document}